\documentclass[twocolumn,amsmath, amssymb, superscriptaddress, pra]{revtex4}
\usepackage{graphicx}
\usepackage{subfigure}
\usepackage{dcolumn}
\usepackage{bm}
\usepackage{epsf}
\usepackage{amssymb}
\DeclareGraphicsExtensions{.eps, .png}
\begin{document}

\author{David S. Simon}
\email[e-mail: ]{simond@bu.edu} \affiliation{Dept. of Physics and Astronomy, Stonehill College, 320 Washington Street, Easton, MA 02357}
\affiliation{Dept. of
Electrical and Computer Engineering \& Photonics Center, Boston University, 8 Saint Mary's St., Boston, MA 02215, USA}
\author{Casey A. Fitzpatrick}
\email[e-mail: ]{cfitz@bu.edu} \affiliation{Dept. of Electrical and Computer Engineering \& Photonics Center, Boston University, 8 Saint Mary's St., Boston, MA
02215, USA}
\author{Shuto Osawa}
\email[e-mail: ]{sosawa@bu.edu} \affiliation{Dept. of Electrical and Computer Engineering \& Photonics Center, Boston University, 8 Saint Mary's St., Boston, MA
02215, USA}
\author{Alexander V. Sergienko}
\email[e-mail: ]{alexserg@bu.edu}
\affiliation{Dept. of Electrical and Computer Engineering \& Photonics Center, Boston
University, 8 Saint Mary's St., Boston, MA 02215, USA}
\affiliation{Dept. of Physics, Boston University, 590 Commonwealth
Ave., Boston, MA 02215, USA}

\begin{abstract}
Recently, a generalization of the standard optical multiport was proposed [Phys. Rev.  A \textbf{93}, 043845 (2016)]. These directionally unbiased multiports
allow photons to reverse direction and exit backwards from the input port, providing a realistic linear-optical scattering vertex for quantum walks on arbitrary
graph structures. Here, it is shown that arrays of these multiports allow the simulation of a range of discrete-time Hamiltonian systems. Examples are described,
including a case where both spatial and internal degrees of freedom are simulated. Because input ports also double as output ports, there is substantial savings
of resources compared to feed-forward networks carrying out the same functions. The simulation is implemented in a scalable manner using only linear optics, and
can be generalized to higher dimensional systems in a straightforward fashion, thus offering a concrete experimentally-achievable implementation of graphical
models of discrete time quantum systems.
\end{abstract}



\title{Quantum Simulation of Discrete-Time Hamiltonians using Directionally-Unbiased Linear Optical Multiports}
\maketitle

\section{Introduction}

Quantum computers have been shown to be capable of performing certain kinds of tasks exponentially faster than classical computers. As a result, an enormous
amount of effort has gone into their development. Although advances have been made, the ultimate goal of a large-scale programable, general-purpose quantum
computer still seems to be a relatively long way off. Therefore it is useful to consider the more easily attainable possibility of special-purpose quantum
computers designed to carry out specific tasks. In particular, one might consider returning to Feynman's original motivation for discussing quantum computers
\cite{feyn}: using simple quantum systems to simulate the behavior of other physical systems.

A number of such quantum simulators appear in the literature; reviews may be found in \cite{aspuru,johnson}. Here we present a new and relatively simple approach
to creating a particular type of quantum simulator using only linear optics. We illustrate the method via the simulation of Hamiltonians for one-dimensional
discrete-time physical models. Such models could represent, for example, the dynamics of spin chains or of electrons hopping along one-dimensional polymers. The
basic approach can be easily generalized to higher dimensions.

The method presented here implements the simulation optically by using chains of simple linear optical units. These basic units  are the directionally-unbiased
optical multiports proposed in \cite{threeport}. These devices are essentially generalized beam splitters, but they differ from the usual beam splitter in
two main respects: (i) they can have any number of input and output ports, and (ii) the photons can reverse direction inside, allowing them to exit back out the
initial input port. These directionally unbiased multiports can be thought of as scattering centers that allow physical implementation of optical quantum walks on graphs \cite{hbf,fh1,fh2,fh3}. (A
different approach to scattering-based walks has also been taken in \cite{tara}.) In a graph model,  an incident photon is constrained at each time step to
scatter into one of a finite number of modes, one of which is the time-reversed version of the input mode. In \cite{threeport}, several applications of these
multiports were demonstrated, including their use as quantum gates for qubits consisting of either single photons or of entangled photon pairs. Here, we show
that they may also be used to implement quantum simulators for Hamiltonians that can exhibit a wide range of behaviors.

In the following sections, we first review unbiased multiports and discrete-time Hamiltonian systems, then give two examples of one-dimensional Hamiltonian
systems whose dynamics can be simulated by quantum walks on sequences of directionally-unbiased three-ports. In each of these cases the same three-port devices
are used, but different Hamiltonians are implemented simply by linking them in different manners. There are a number of obvious generalizations that can be made,
such as using $n$-ports with $n>3$, of allowing the properties of the $n$-ports to vary with position, or of connecting the $n$-ports into two- and
three-dimensional networks. But even in the simple cases considered here (one-dimensional three-port chains with all of the three-ports having identical
parameters), an array of different physical system behaviors can already be seen to occur.  The examples discussed show that systems with both spatial and
internal degrees of freedom can be simulated, and it is clear that, by means of multiports with large $n$, high-dimensional systems can be implemented in a
straightforward manner by simply extrapolating the same methods. Since there is no impediment to putting each linear-optical multiport on a chip for high
stability, the approach is highly scalable.

In addition to the topologically-trivial examples discussed in the current paper, it will be shown elsewhere \cite{simtop} that a different arrangement of the
same multiports can simulate physical systems that support states with non-zero winding number and topologically-protected boundary states.

\section{Directionally-Unbiased Multiports}\label{unbiased}

\subsection{Directionally-unbiased multiports and photon reversibility} 
\begin{figure}
\begin{center}
\subfigure{
\includegraphics[scale=.20]{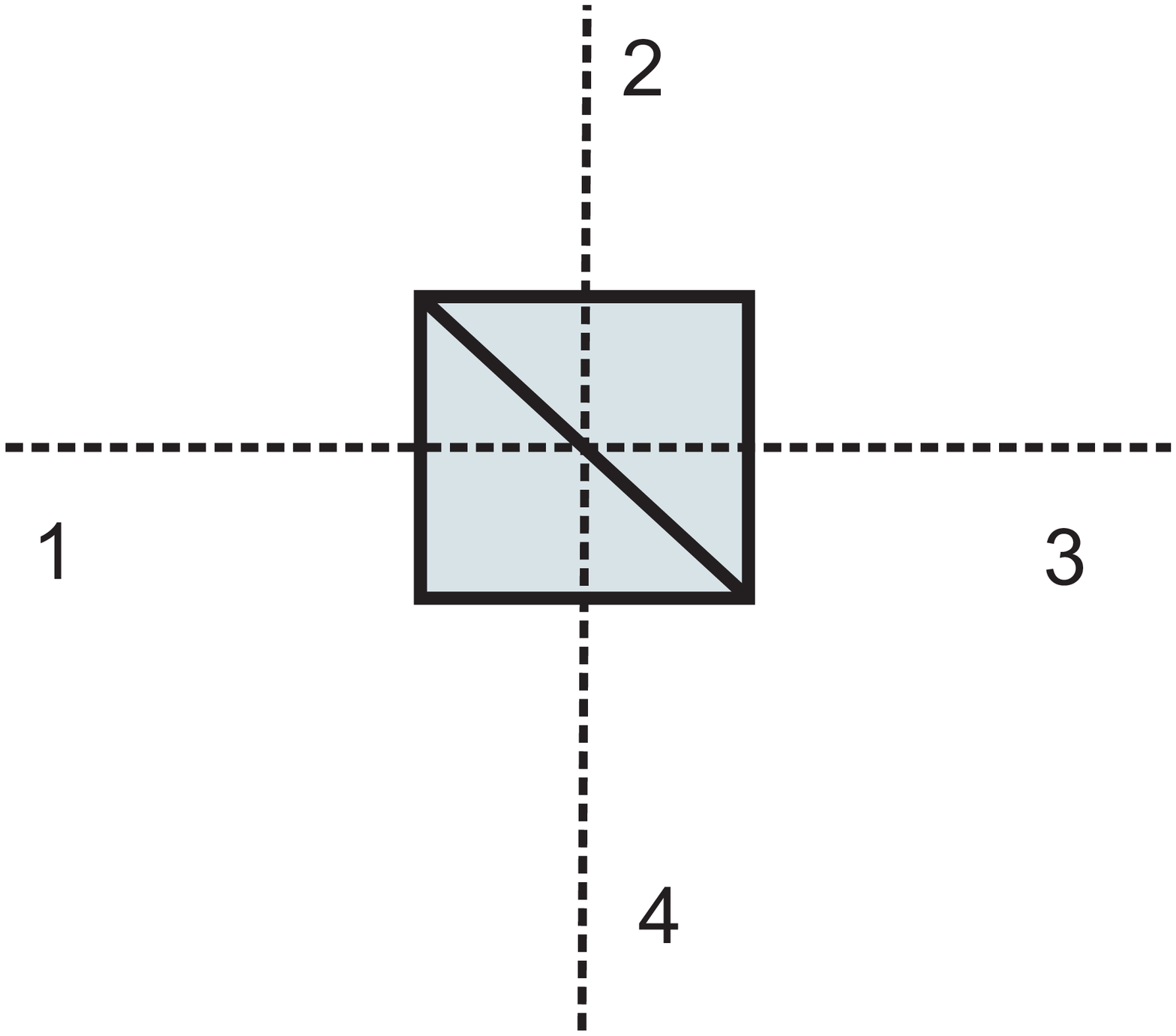}}
\qquad \qquad \subfigure{
\includegraphics[scale=.20]{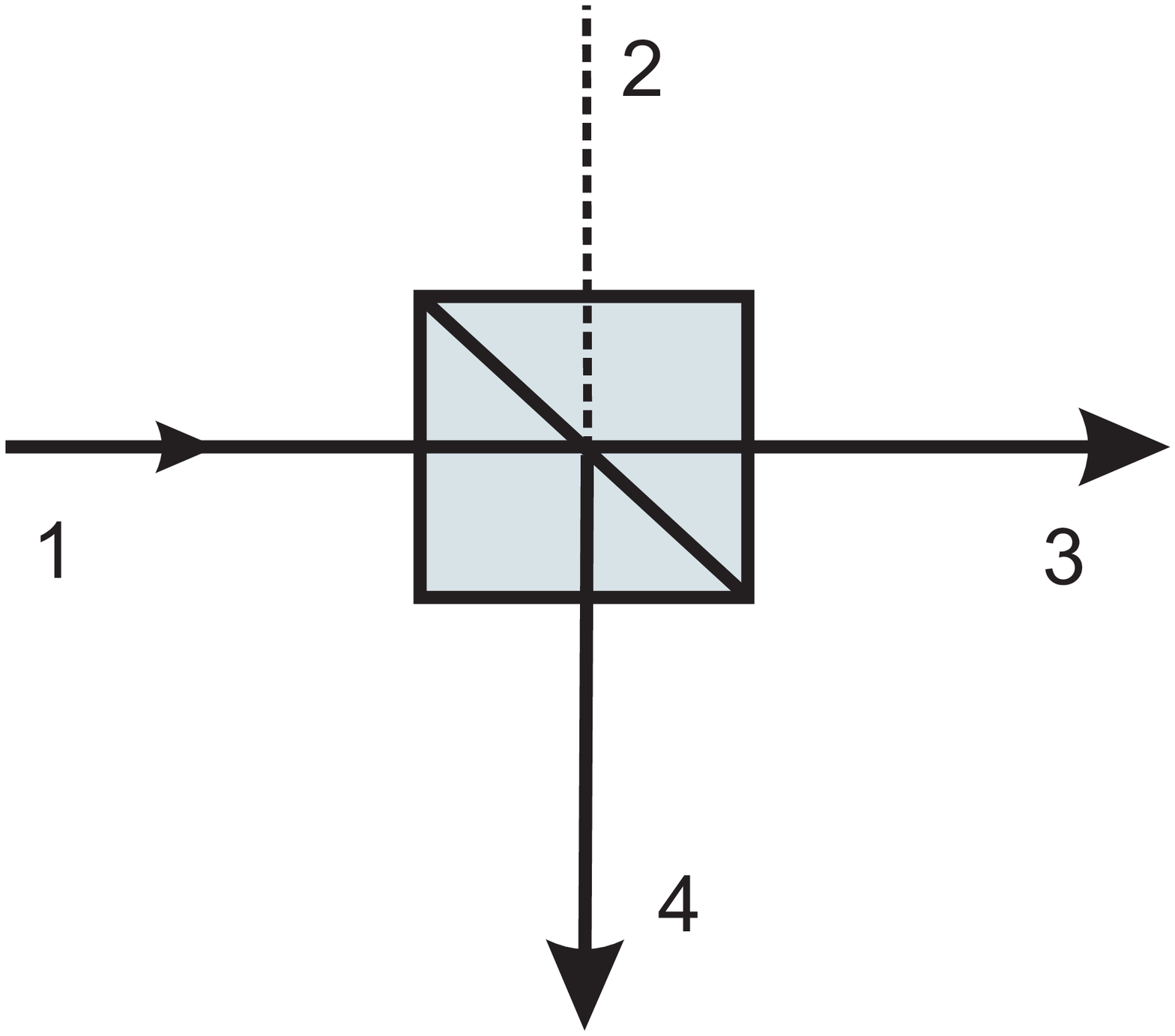}}
\caption{(Color online) (a) The beam splitter is a four-port device. Initially, all four ports are symmetric. (b) But when a photon is input to one port the symmetry is broken: the photon can exit from ports $3$ and $4$, but not $1$ or $2$. The fact that the photon cannot reverse direction and exit out input port $1$ is referred to by saying that the beam splitter is \emph{directionally biased}: the presence of the photon in one port biases the output toward two of the four possible output directions.  }\label{bsfig}
\end{center}
\end{figure}

An ordinary beam splitter or its standard multiport generalization is, in a certain sense, a
one-way device. Although all four ports of a beam splitter are initially on an equal footing, once a photon enters one port (say port $1$ of Fig. \ref{bsfig}),
then it can exit only from ports $3$ or $4$, not from $1$ or $2$. However it is possible to construct a multiport system, with any number $n$ of ports, such that
a photon entering any port can leave any port. In particular, the photon may exit back out the initial input port. Such a multiport, which allows the light to
reverse direction, is referred to as \emph{directionally-unbiased}.

\begin{figure}
\begin{center}
\subfigure{
\includegraphics[scale=.2]{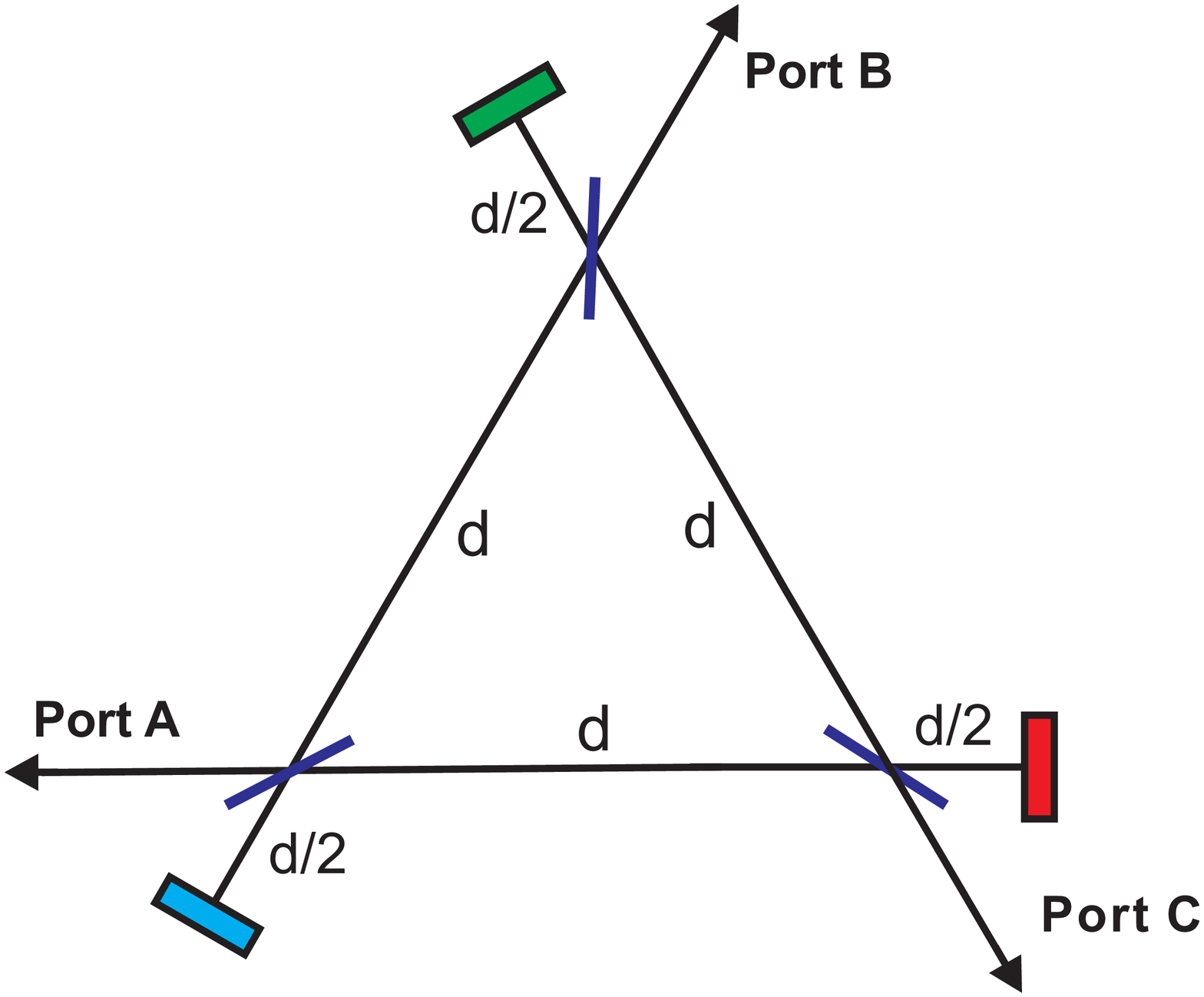}}
\quad  \subfigure{
\includegraphics[scale=.2]{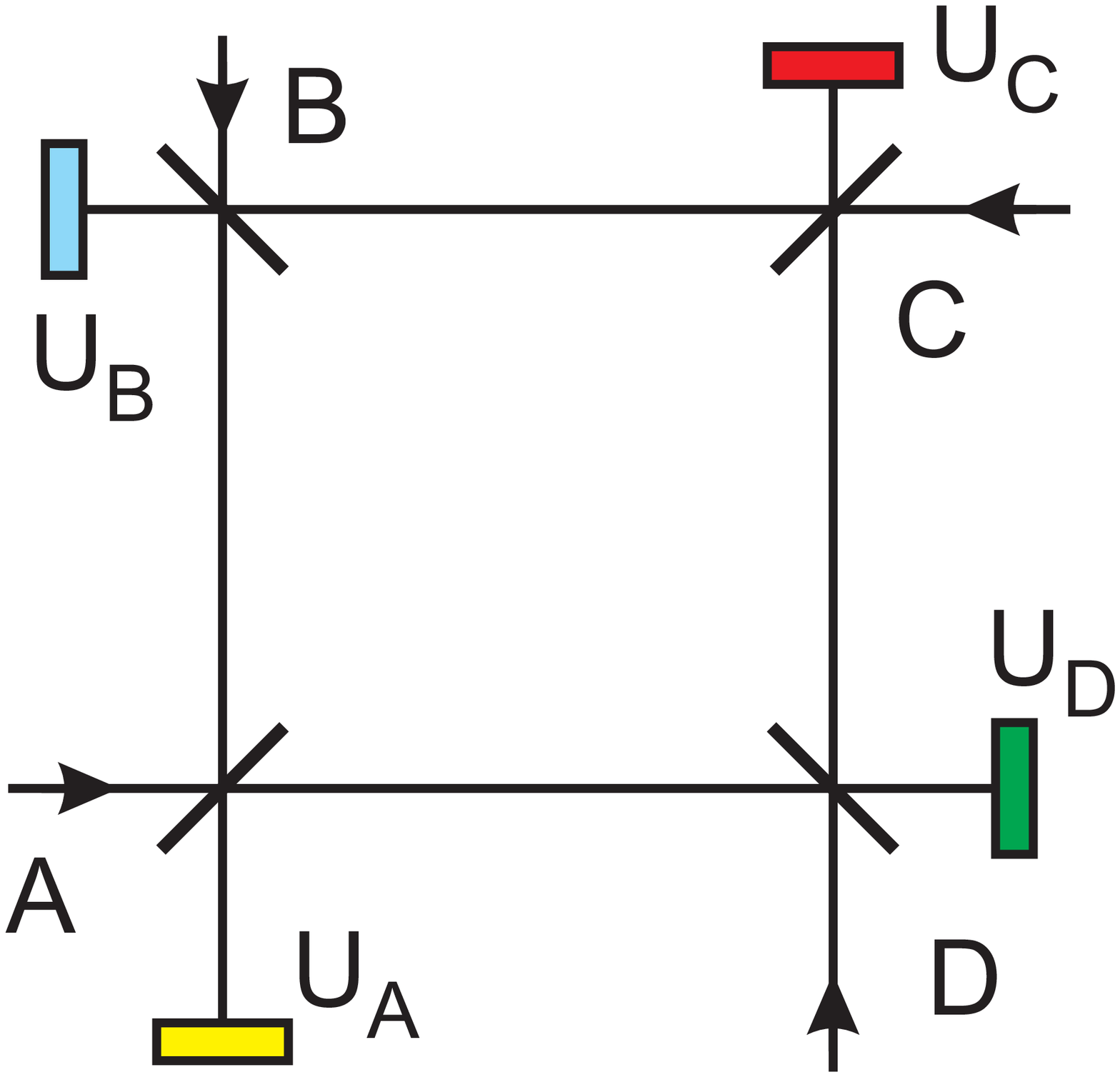}}
\quad  \subfigure{
\includegraphics[scale=.2]{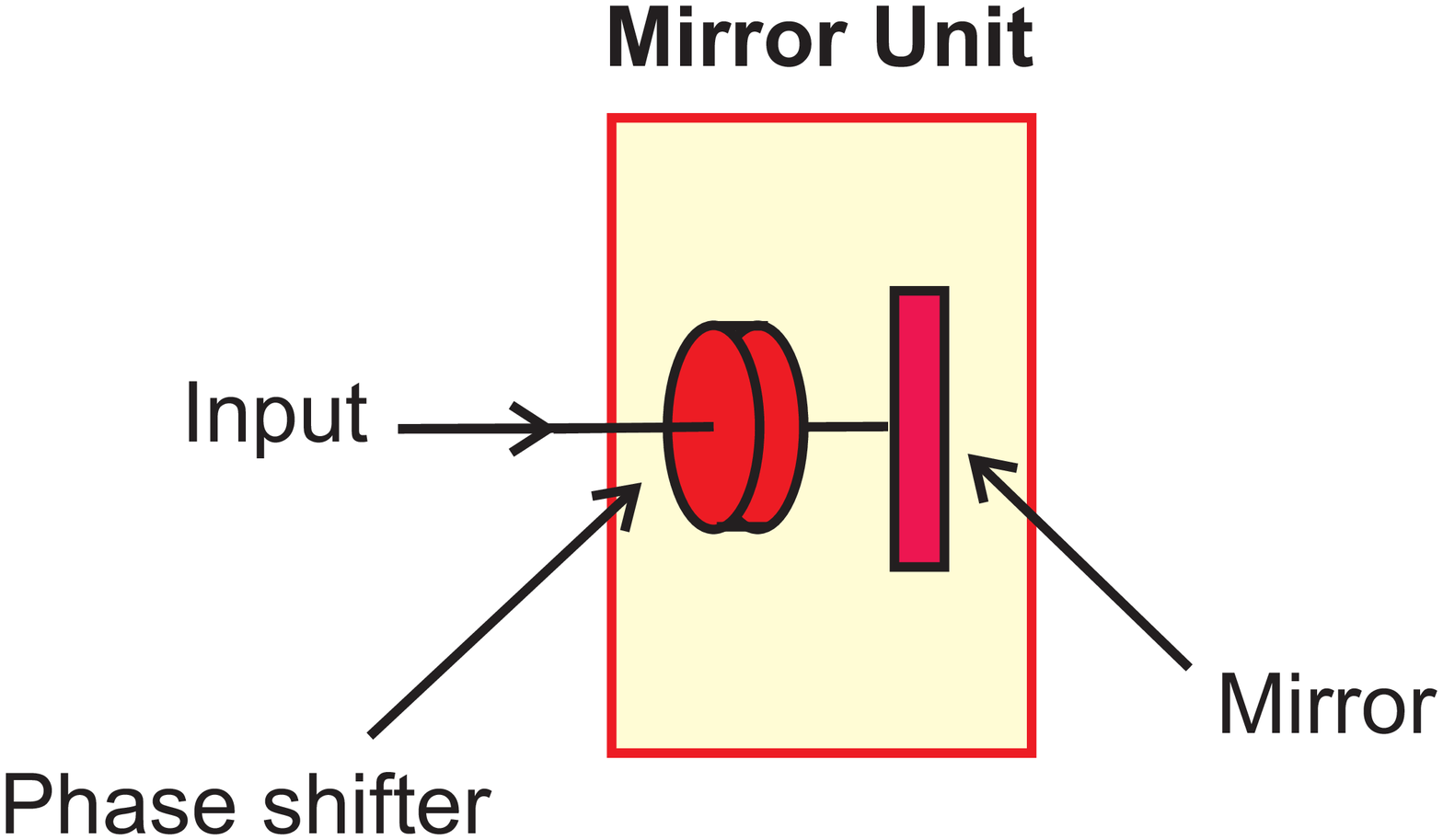}}
\caption{(Color online) (a) The directionally-unbiased three-port. (b) The directionally-unbiased four-port. The rectangles after the beam splitters in
(a) and (b) represent the vertex unit shown in (c); this unit consists of a mirror and a phase-shifter. The distance between each beam
splitter and the adjacent mirror unit is half the distance $d$ between one beam splitter and the next. }\label{nportfig}
\end{center}
\end{figure}

Examples of unbiased $n$-ports for $n=3$ and $n=4$ are shown in Fig. \ref{nportfig}(a) and (b). As in \cite{threeport} the focus here will be on the three-port
device. The cases of higher $n$ are similar. The key to constructing such directionally unbiased multiports using only linear optics is to build it from vertex
units of the form shown in Fig. \ref{nportfig}(c). Each unit is at one port of an ordinary beam splitter, and is constructed from a mirror and a phase shifter.
At each beam splitter, one port is used for input/output to the device, and two ports feed into the interior of the multiport. The remaining beam splitter port
reflects back on itself via the mirror inside the vertex unit. The beam splitter-to-mirror distance ${d\over 2}$ is half of the distance $d$ between the vertex
units in the multiport. The internal phase shifter provides control of the properties of the multiport, since different choices of phase shift at the vertices affect how
the different photon paths through the device interfere with each other.

Given any input and any output port, there will be multiple paths joining them, and these paths will have different lengths. Therefore, for any input there will
be transient states within the device that will decay over time as longer paths exit the multiport. Because of this range of exit times, there will be a
lengthening of input pulses, and a gradual loss of coherence. These effects may be minimized by taking the physical size of the multiports small enough compared
to the distances between them, and by taking the pulses to be sufficiently short compared to the other time scales in the system. Quantitative estimates of the
sizes and time scales required may be found in \cite{threeport}; here we simply assume that the multiport is small enough to be treated as coherent and point-like. In that
case, the three-port takes an input state $|\psi_0\rangle$ to an output state $|\psi\rangle =U|\psi_0\rangle$, where
\begin{equation}U=-{i\over 3}\left( \begin{array}{ccc} 1 & -2 & -2\\ -2 & 1 & -2\\ -2 & -2& 1
\end{array}\right) .\label{U3port}\end{equation} Here the rows and columns refer to the three ports $A$, $B$, $C$. This matrix reproduces the scattering amplitudes of the
diamond-shaped graph used as an example in \cite{fh1,fh2}.

For an arbitrary $n$-port with the same choice of vertex phases, the matrix of Eq. \ref{U3port} generalizes to
\begin{equation}U_n=-{i\over n}\left( \begin{array}{cccc} n-2 & -2 & -2 & \dots\\ -2 & n-2 & -2\\ -2 & -2& n-2 \\ \vdots & & & \ddots
\end{array}\right) .\label{Unport}\end{equation} Note that this is the form of an $n$-dimensional Grover coin \cite{carn}, and so provides a
physical implementation of the Grover coin for quantum walks in any number of dimensions. The behavior of this matrix versus $n$ should also be noted: for $n=3$
the exit probability is lowest at the input port. For $n=4$, the exit probabilities at all four ports are equal, while for $n>4$ the exit probability becomes
highest at the input port. As $n\to \infty$ the multiport acts effectively as a mirror, with the probability of exiting back out the input port going to
$100\%$.

\subsection{Strictly unbiased multiports with equal probabilities} For the choice of vertex phases given above, it is clear that the exit probability
at the input port differs in general from the probabilities to exit at the other two ports. In the three-port case, for example, input at exit $A$ leads to exit
probabilities $P_A={1\over 9}$ and $P_B=P_C={4\over 9}$. The term ``directionally-unbiased'' here refers to the fact that both forward and backward are possible,
not necessarily that all exit probabilities are equal. However, for some values of phases the multiports \emph{are} unbiased in the stricter sense of having
equal exit probabilities at all ports. As can be seen from Eq. \ref{Unport}, this is already true for the four port with the choice of phases used here: the exit
amplitudes at all four ports are the same, up to a minus sign.

The same equality of exit probabilities can also be achieved for the three-port by appropriate
choices of vertex phases; for example, for $\phi_A=\phi_B=\phi_C={\pi \over 6}$ all
three ports have exit probability $1\over 3$, with transition matrix \begin{equation}U={1\over \sqrt{3}}e^{2\pi i\over 3}\left( \begin{array}{ccc} e^{-2\pi i\over 3} & 1 & 1\\
1 & e^{-2\pi i\over 3} & 1\\ 1 & 1& e^{-2\pi i\over 3}
\end{array}\right) .\end{equation} Such cases where the exit probabilities are all equal will be referred to as \emph{strictly unbiased}. The strictly unbiased multiport is useful for some applications, such as the conversion of position eigenstates into momentum eigenstates (see Section \ref{preparesection}) or for quantum state discrimination methods (to be discussed elsewhere).

In the remainder of the paper we will not require this strict unbiasedness. We will instead follow the example of \cite{threeport}, focusing on the special case of the
three-port with equal phases of $\phi =-{{3\pi}\over 4}$, leading to the transition matrix of Eq. \ref{U3port}. This choice is
convenient for quantum walk applications because all of the transition amplitudes have the same phase (up to minus signs), reducing the number of phase factors that have to be tracked and eliminating
complicated interference terms.

\section{Discrete-Time Hamiltonians}\label{discsection}

In a discrete time system, the Hamiltonian is obtained from a discrete-time evolution matrix $U$ that takes the system forward one time-step. So if the
initial state is $|\psi(0)\rangle$ and the unit time step is $T$, then the state at time $nT$ is
\begin{equation}|\psi(nT)\rangle =U^n |\psi(0)\rangle .\end{equation} The evolution operator can then be written (for $\hbar =1$) in the form $U=e^{-i\hat HT},$ which defines the
discrete time Hamiltonian, $\hat H$. The Hamiltonian generates time evolution, and the matrix elements of $U$ give the transition amplitudes per time step between the states  of the system.

If the system is spatially periodic, then Bloch's theorem says that the solutions should be of the form \begin{equation} \psi (x) \sim u(x)
e^{ikx},\end{equation} where $u(x)$ is a periodic solution and $x$ is the spatial position. The period of $u(x)$ is the same as that of the underlying system.
The phase factor  $e^{ikx}$  defines the crystal momentum or quasi-momentum $k$, which describes how fast the phase accumulates as you move along the periodic
lattice. Because of this phase factor, the periodicity of $\psi (x)$ is not necessarily the same as the periodicity of the lattice. In addition, a quasi-energy $E$ can then be defined, which will be possibly multi-valued function of $k$. Each steady state of the system is characterized by a fixed value of $k$ and $E$.
%

The role of position in the following will be taken by the dimensionless integer $m$ that labels the lattice sites, where each lattice site consists of some combination of unbiased multiports. Because the position is given by a dimensionless variable, the quasi-momentum will also be dimensionless. A single Brillouin zone runs from $0$ to $2\pi$, and $k$ is only conserved modulo $2\pi$.

The Hamiltonian generates time evolution in some space that may include both spatial and internal degrees of freedom.  As the momentum is varied over the width
of a full Brillouin zone, the Hamiltonian will trace out a closed path in this internal space. There may be topological obstructions that prevent some of these
paths from being contracted to a point or from being deformed continuously into each other as the system parameters are varied. When such obstructions exist,
then the paths fall into different topological classes, categorized by the values of topologically invariant quantities such as the winding number in
one-dimensional lattice systems or the Chern number in two-dimensional systems. When two systems with different topological invariants are brought into contact,
consistency between the solutions in the two subsystems requires that the band gap between quasi-energy levels vanish at the boundary. The closing of the gap
allows states to change winding number, and implies the existence of states that are exponentially localized in the vicinity of the boundary
\cite{kitagawa,asboth}. In this paper we restrict ourselves to states with vanishing winding number; the extension to systems with non-trivial topological
aspects will be examined elsewhere. Reviews of topological insulators and related ideas may be found in \cite{hasan,asboth}.

If the Hamiltonian describes internal, as well as spatial, degrees of freedom, then it is convenient to write it as a position-dependent matrix in
the space of internal variables. For example, if the internal space is two-dimensional, as is the case for electron spins or photon polarizations, then the
Hamiltonian will generically be of the form
\begin{eqnarray}\hat H(k) &=& d_0(k) I + d_x(k)\sigma_x+d_y(k)\sigma_y +d_z(k)\sigma_z \\ &=& d_0(k) I +\bm d(k)\cdot \bm\sigma ,\end{eqnarray} where the
$d_j(\bm x)$ are real functions. The Pauli matrices appear are the generators of $su(2)$, the Lie algebra of traceless $2\times 2$ unitary matrices. Any $2\times
2$ unitary matrix, such as the matrices representing single qubit quantum gates, can be built from a superposition of these matrices and the identity, so the
Hamiltonian describing the dynamics of the two-dimensional internal space will be formed from them. Similarly, in the example to be discussed in Section
\ref{internalsection}, the Hamiltonian may be expressed as a matrix in a three-dimensional internal space, analogous to the state space of a spin-1 particle or
of a quark in its three-dimensional color space. In this case, $\hat H$ can be written in the form
\begin{equation}\hat H= d_0 I +\sum_{j=1^8}d_j\Lambda_j = d_0 I +\bm d \cdot \bm \Lambda,\end{equation} where the $3\times 3$ Gell-Mann matrices $\Lambda_1\dots \Lambda_8$, which are widely used
in elementary particle physics, form a basis for the algebra of $3\times 3$ traceless, unitary matrices, $su(3)$. As $k$ is varied, the Hamiltonian will trace
out a path in the resulting space.

In the following sections we show that chains of directionally-unbiased three-ports allow simulation of two different Hamiltonians with very different
characteristics. In Section \ref{3pointsection} a system with a single three-valued degree of freedom will be presented; these three values may be thought of as
either three positions on a periodic spatial lattice, or as three values of an internal variable at a single, fixed spatial point. The second example, in Section
\ref{internalsection} will allow simulation of both spatial and internal degrees of freedom simultaneously.

\section{Dynamics on a Discrete Three-Point Configuration Space}\label{3pointsection}

\subsection{A discrete-time system} Recall that the unbiased three-port is described by the matrix $U$ of Eq. \ref{U3port}, where the rows and columns refer to the
three ports $A$, $B$, $C$. For convenience in what follows, let us instead label the three ports by numbers (modulo 3) instead of letters: $$ A=1 \mbox{ (mod
3)}, \quad B=2 \mbox{ (mod 3)}, \quad C=3 \mbox{ (mod 3)}.$$ The three input/output ports are now thought of as discrete points on a circle; with each passage
through a multiport, photons hop among these points.  $U$ has three eigenstates, given (up to an arbitrary overall phase) by
\begin{eqnarray} & & |\psi_1\rangle ={1\over \sqrt{3}}\left( \begin{array}{c} 1 \\ 1 \\ 1 \end{array}\right), \quad
|\psi_2\rangle ={1\over \sqrt{2}}\left( \begin{array}{c} -1 \\ 0 \\ 1 \end{array}\right), \\ 
&\; & |\psi_3\rangle ={1\over \sqrt{2}}\left( \begin{array}{c} -1 \\ 1 \\ 0 \end{array}\right),
\end{eqnarray} with respective eigenvalues \begin{equation}\lambda_1 =+i, \qquad \lambda_2=\lambda_3 =-i.\label{eigenstates}\end{equation}


\begin{figure}
\begin{center}
\subfigure[]{
\includegraphics[scale=.25]{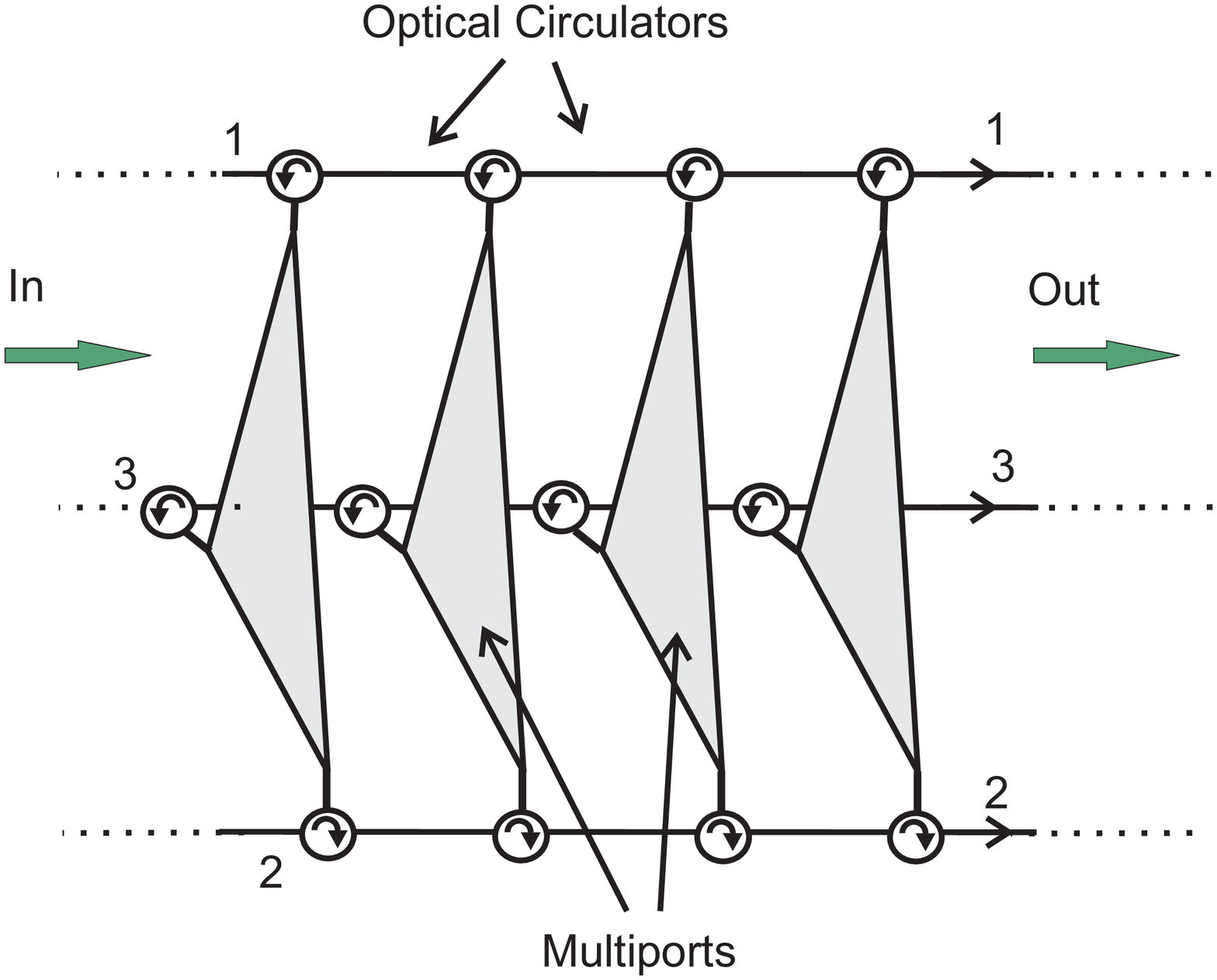}}\label{simplefig}\qquad\qquad
\subfigure[]{
\includegraphics[scale=.25]{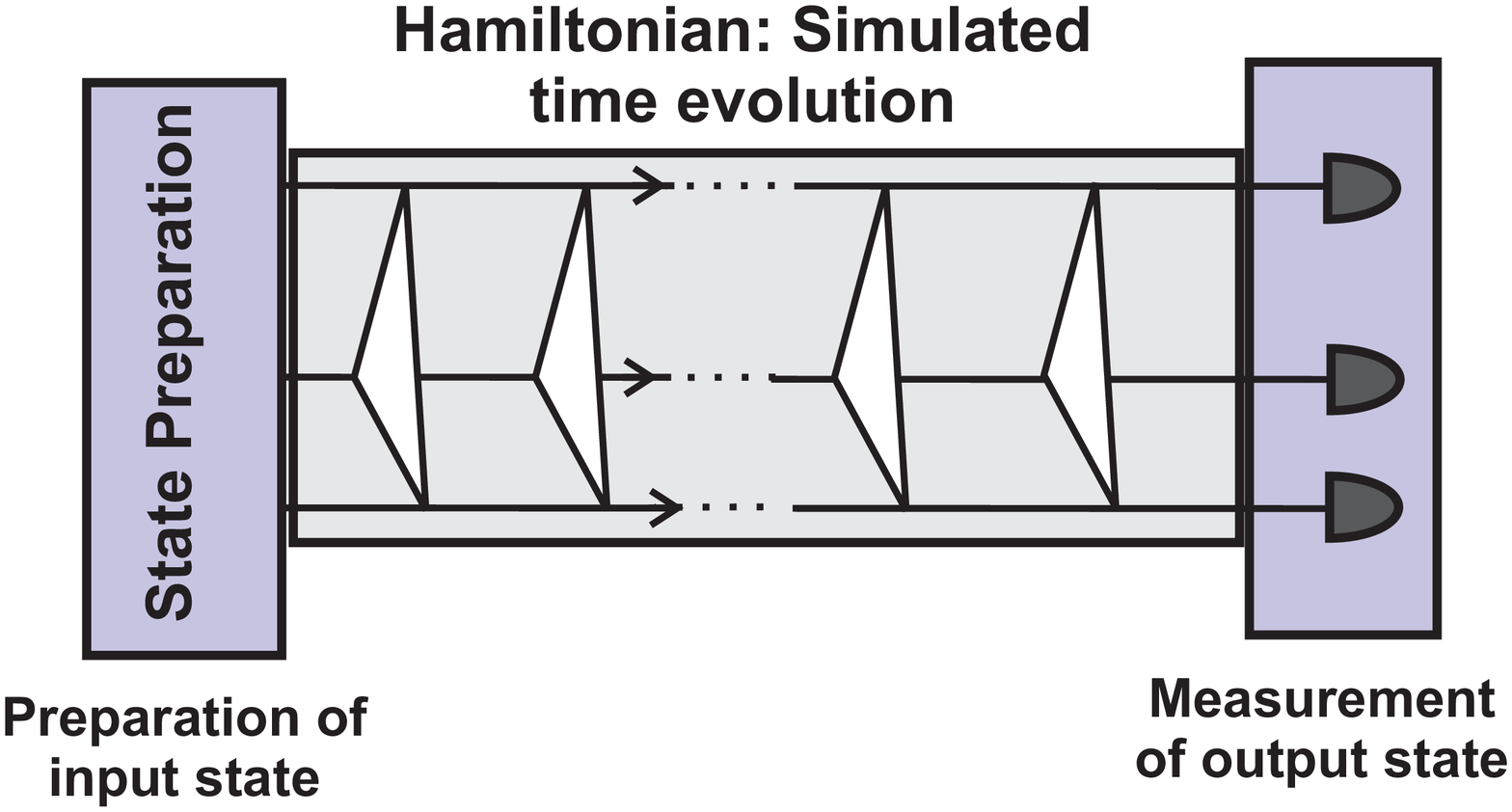}}\label{simplefiga}
\caption{(Color online) (a) A string of multiports connected sequentially, with each output of one feeding directly into the input of the next. There is a
three-state quantum walk among the three lines as the photons progress spatially from left to right. (For clarity, the figures are not drawn to scale: the multiports should be very small compared to the distances between them.) (b) State preparation occurs on the left and photons are coupled into the system through the three input ports. Detection occurs on the right, with a detector in each line. Depending on the experiment being performed, the detectors may be connected in coincidence.}\label{simfig}
\end{center}
\end{figure}

Now suppose a sequence of three-port units are connected as shown  in Figure \ref{simfig}(a).
Each output of one multiport feeds directly as input to the next. It can be arranged (for example by means of optical circulators) so that the photons always travel from left to right and do not reflect backward. In other words, the spatial location increases monotonically and plays the role of time. It is assumed that the multiports are very small compared to the length of the lines joining them, so that the time spent \emph{inside} the multiport can be taken to be negligible.
If photons are fed in from the left, they can be thought of as experiencing a periodic
potential, as in the lattice of a one-dimensional solid or a one-dimensional polymer. Each multiport is viewed as the site of an ``interaction'', where the state of the photon can change. Over $n$ time steps the unitary transition matrix $U$ is simply applied $n$ times.

\subsection{Position-space Hamiltonian} Given that $U$ is known, the Hermitian Hamiltonian operator $\hat H$, is easily found:
\begin{equation}\hat H =+i\ln U={\pi\over 6} \left( \begin{array}{ccc} 1 & -2 & -2\\ -2 & 1 & -2\\ -2 & -2& 1
\end{array}\right) =i{\pi\over 2}U.\label{HandU}\end{equation}  Here, the discrete unit of time is taken as the time to go from one multiport to the next; in other words, units are chosen so that $T=1$.
For this system $\hat H$ turns out to be proportional to $U$ itself, so the eigenstates are those given in Eq. \ref{eigenstates},
with corresponding energy eigenvalues \begin{equation}E_1=-{\pi\over 2}, \qquad E_2=E_3=+{\pi\over 2}. \end{equation} These eigenvalues can also be found by computing the expectation value of the Hamiltonian in each of the three
eigenstates, $E_j=\langle\psi_j |\hat H |\psi_l\rangle $ for $j=1,2,3$. In terms of the position eigenstates $|m\rangle$, the Hamiltonian operator can be written as \begin{equation}\hat H={\pi\over 6}\sum_{m=1}^3 \left[ |m\rangle\langle m| -2\left( |m+1\rangle\langle m| +|m\rangle\langle m+1| \right) \right] .\end{equation}
Note that the eigenvalues of this Hamiltonian will not be the photon energies $\hbar \nu$, but instead give the "quasi-energies" associated with the various wave solutions propagating in the system. Higher quasi-energy occurs for standing waves that oscillate more rapidly in space.

In this Hamiltonian, the first term is an effective mass term, giving the state a sort of ``inertia'' or probability of being at the same point at successive
times. It contributes a constant, position-independent background energy $\pi\over 6$. The other terms serve as a nearest-neighbor interaction. This system can
be viewed in several different ways. It can be thought of as a single three-state system, where the three links or positions labeled by $m$ correspond to the
three states. Alternatively, this may be seen as a coupled set of three two-level systems: each of the three links can be in the ground state (no photon) or an
excited state (one photon).

\subsection{Momentum-space Hamiltonian} The quasi-momentum eigenstates $|k\rangle$ are found by Fourier transforming the position states $|m\rangle$:
\begin{equation}|k\rangle ={1\over \sqrt{3}} \sum_{m=1}^3 e^{imk}|m\rangle .\label{kstate}\end{equation} In momentum space, the Hamiltonian has the form
\begin{equation}\hat H(k) =  {\pi\over 6}\sum_{k}  \left(1-4\cos\left(k\right)\right) \cdot |k\rangle\langle k|. \end{equation}  For a three-dimensional configuration space there will be three momentum eigenstates, with momenta $k_n={{2\pi n}\over 3}$ for $n=0,\pm 1$; these have the energies given above, and any other state will be a superposition of them. When a single photon is input into the system, the physical meaning of the momentum eigenstates is easy to identify:

(1) There is a spatially uniform ``rest'' state with momentum $k=0$ and energy $E_1=-{\pi\over 2}$. In this state, the photon amplitude is evenly spread among the three ports. This state is constant in time.

(2) There is a steady state with the average photon position moving clockwise by one step per unit time, $\Delta m =+1$. This state has momentum $k={{2\pi}\over 3}$ and energy $E_2=+{\pi\over 2}$.

(3) The remaining steady state has the average photon position moving counterclockwise by one step per unit time, $\Delta m =-1$. This state has momentum $k=-{{2\pi}\over 3}$ and energy $E_3=+{\pi\over 2}$.

Because of the periodic nature of the configuration space, these are the only possibilities with fixed $k$. For example, it can be seen that the motion with $\Delta m=-2$
per step is the same as that with $\Delta m=+1$ per step. All other periodic motions around the triangle are therefore equivalent to one of the three cases above.

In the preceding, $m$ has represented a spatial degree of freedom, but we may also think of it as simulating an internal degree of freedom at a single point.
As one possible application of this approach, the three ports may be thought of as representing the three spin states of a spin-$1$ particle.
Although the vertices of the multiports have been held at fixed parameter values in this paper, the mirror reflectances and the phase shifts added at each vertex
can be changed, which allows control over the interaction terms and hopping amplitudes; this could for example lead to simulation of the behavior of a spin $1$
particle in a time-dependent potential. By replacing the unbiased three-port with a larger $n$-port, the same procedure can be used to model dynamics in an
$n$-dimensional internal configuration space at each lattice point, or in other words a system of spin $j$, with $2j+1=n$.

\subsection{State preparation and measurement}\label{preparesection}

The system of Fig. \ref{simfig}(a) simulates the time evolution and the Hamiltonian. To complete the description of the system, the initial state preparation and the measurement of the final state must be included. As shown in \ref{simfig}(b), the input state is introduced at left, the sequence of multiports simulates time evolution as the photons move left to right, and measurements are made at the final output ports on the right.

To introduce an initial position state, $|m\rangle$, single photon states are inserted at one of the left-hand input ports. High quality approximations to single photon input states may be achieved by means of highly attenuated laser coherent states or via heralded parametric down conversion. The three inputs may be populated randomly by a sequence of such states in order to study the full dynamics of the system.

Arrangements of beam splitters also allow a single photon state to be uniformly superposed over the three input ports. In fact, an additional three port may
be used to prepare such a superposition state: as mentioned in Section \ref{unbiased}, using phase shifts of $\pi\over 6$ at the vertices of a three-port
makes the exit amplitudes of all three ports equal (up to phase) for any input port. By addition of appropriate external phase shifts to the three output lines, momentum eigenstates of the form of Eq. \ref{kstate} may be initialized from this superposition.

At the right end of the system, a single-photon detector may be placed at each output line. The final state for a given input is then built up by making multiple
measurements to determine the various transition probabilities. The measurements may be taken one step further by using homodyne detection to determine the
output phases relative to the input. In this way, the transition amplitudes, not just the probabilities, may be determined.

In the current paper, we consider only separable input states, but various forms of superposition states and entangled states could also be considered. In this
case, the detectors would be connected in coincidence in order to determine the final state correlations.

In passing, it should be pointed out that the method described above for producing momentum eigenstates also provides a simple means of producing spatial W-states. With $\pi\over 6$ vertex phases and appropriate phase shifts at the output lines, a three-port with a single photon input at any vertex can be made to output the state ${1\over \sqrt{3}}\left( |100\rangle +|010\rangle +|001\rangle\right) ,$ with equal amplitudes to be in three different spatial modes. Here, $|mnp\rangle$ denotes the state with $m$, $n$, and $p$ photons exiting respectively at ports $A$, $B$, and $C$. Clearly, this tripartite W-state can be generalized to an n-partite W-state by replacing the three-port with an $n$-port of larger $n$.

\begin{figure}
\begin{center}
\subfigure[]{
\includegraphics[scale=.30]{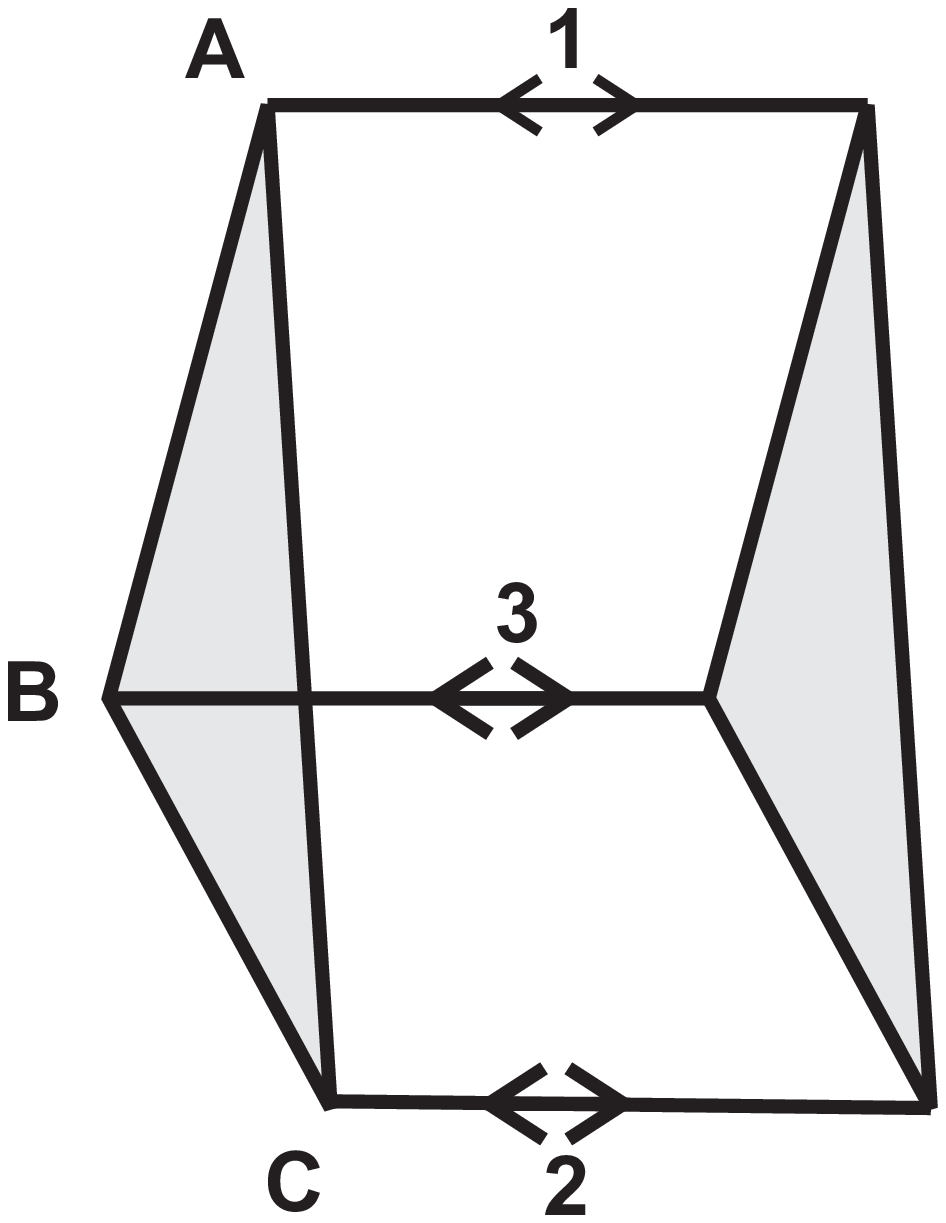}}\qquad\qquad
\subfigure[]{
\includegraphics[scale=.25]{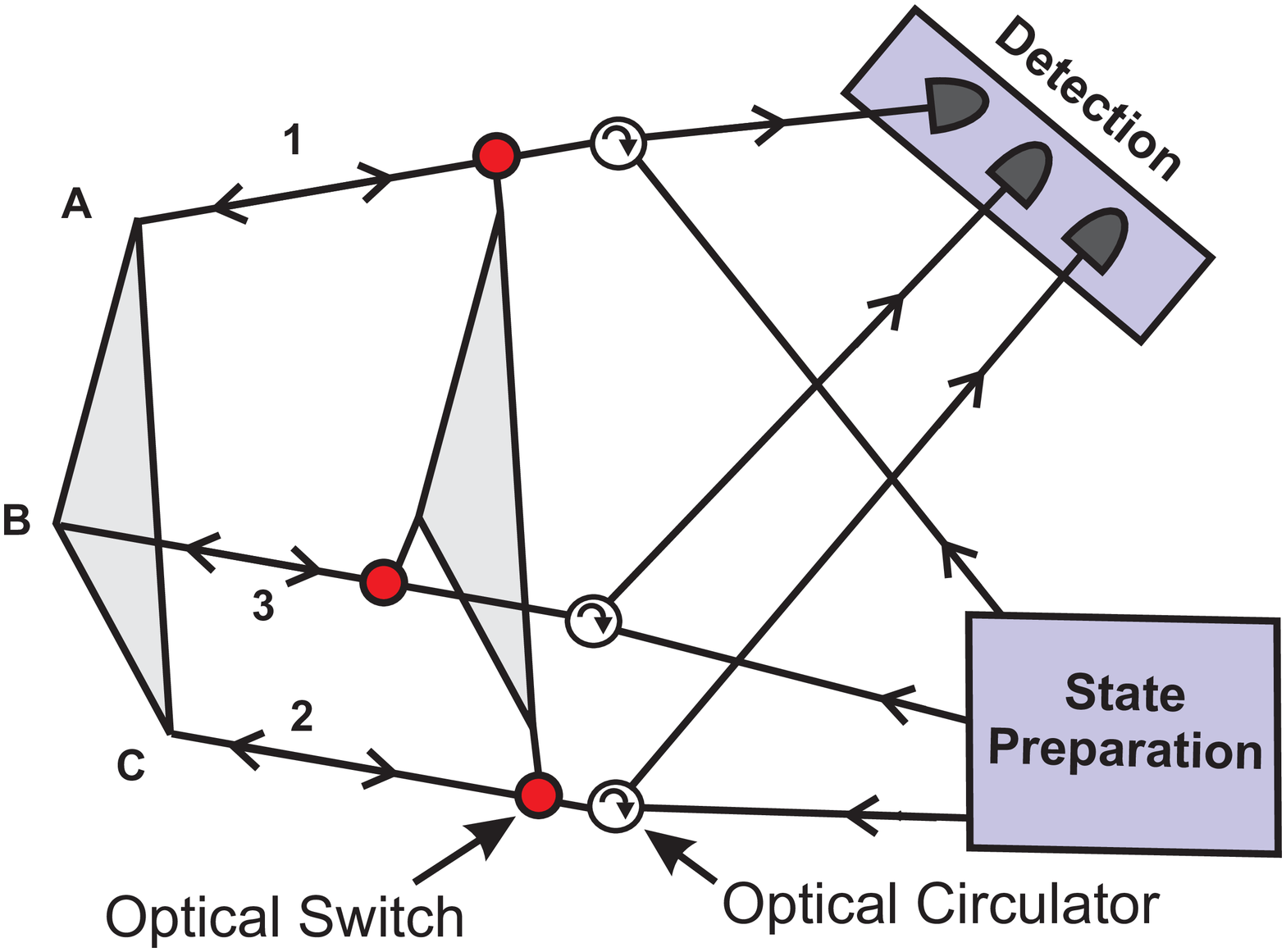}}
\caption{(Color online) (a) A compact system equivalent to that of Fig. \ref{simfig}, but which operates using just two multiports; the spatial evolution from left to right in  Fig. \ref{simfig} is replaced by time evolution within a single region. The photons travel in opposite directions on successive time steps. (For clarity, the figures are not drawn to scale: the multiports should be very small compared to the distances between them.) (b) After coupling the input into the system, a gated optical switch allows measurement of the output state after a fixed number of time steps. }\label{compactfig}
\end{center}
\end{figure}

\subsection{A compactified system} Instead of using a chain of multiports, an equivalent system can be formed with fewer resources by taking just two multiports connected as shown Fig. \ref{compactfig}(a). The photons then reflect back and forth
between the two multiports, with no circulators needed. At each discrete time step the photons will alternate direction, but will always remain localized in the same spatial
region, between the two multiports. The time evolution then generates superpositions of random sequences of values $m_1,m_2,m_3,\dots$, where each value in the sequence is given
by the label of one of the three ports. Thinking of the lines between the multiports as three spatial positions,  the Hamiltonian generates dynamics that
cause random motions among these three discrete spatial points. There will then be a momentum variable (a dimensionless quasi-momentum) describing the
motion between these points. There can be superpositions between clockwise and
counterclockwise motions, allowing standing waves to occur.

Optical circulators are used to separate input and output, with electro-optical switches used to release the output to the detectors after the desired number of time steps as shown in Fig. \ref{compactfig}(b). In this way, walks of arbitrarily long duration can be simulated with a single unit containing only two multiports. The maximum number of steps achievable will then be limited only by coherence considerations.

\section{A System with Both Spatial and ``Internal'' Degree of Freedom }\label{internalsection}

\subsection{Time-reversible systems} The previous section allowed simulation of a system with one three-valued degree of freedom, which could be taken to represent either a spatial or internal degree of freedom. The system of this section incorporates both spatial and internal variables  simultaneously.

An additional important difference between this section and the previous one is related to the idea of reversibility. In the last section it was assumed that the system had been arranged to make sure that the photons flowed only in one direction, from left to right. The states were labelled by the positions of the photons, with no need to distinguish direction of motion. In this case, the unitary transition matrix was found to be proportional to a Hermitian matrix that could serve as the Hamiltonian.

Now we wish to restore the time reversal symmetry of the system and allow the light to move in any direction though the network. Therefore, we need to
distinguish states moving in different directions, and for any possible transition that can occur we must also allow its time-reversed version. The time-reversed
transition matrix is obtained from $U$ by taking its Hermitian conjugate, $U^\dagger$. If the states are labelled by both position and direction of motion of the
photons, then the number of possible states is doubled. Imagine that the states of the three-port are now labelled by both the port label and a binary label such
as ``ingoing/outgoing'' or ``left-moving/right-moving''. Then the matrix $U$ is replaced by the larger matrix \begin{equation} \hat H=\left( \begin{array}{cc}0&
U^\dagger\\ U & 0\end{array}\right),\label{Hreversed}\end{equation} which includes the time-reversed transitions.  One of the nonzero blocks represents
transitions from ingoing or left-going states at a vertex to outgoing or right-going states; the other block represents transitions in the other direction.
Because $U$ is unitary, $U^\dagger U=1$, it is easily checked that this expanded matrix is both unitary and Hermitian and that it squares to the identity, $\hat
H^2=I$. As a result, we find that the corresponding matrix \begin{equation}V\equiv e^{-i\hat HT} = I\cos T -i\hat H\sin T\label{TandV}\end{equation} is also
unitary. Thus, from the transition amplitudes determined by the original $U$, we form (i) a new operator $\hat H$ which can serve as a Hamiltonian, and (ii) a
corresponding unitary operator $V$ that acts as the full transition operator associated with this Hamiltonian. $V$ is double the dimension of the original $U$.
When $\hat H$ is of the form of Eq. \ref{Hreversed}, notice that if the period is taken to be $T={\pi\over 2}$ then the Hamiltonian and the transition matrix $V$
are the same, up to overall constants, as was the case in the previous section.

Additional terms may also appear in the diagonal blocks of $\hat H$ representing, for example, left $\to $ left or right $\to $ right transitions, as long as
they are real and diagonal or occur in Hermitian conjugate pairs. Such extra terms in the diagonal block can appear because of, for example, the possibility of a
left-moving mode on one edge becoming left-moving on an adjacent edge at the next step.


\subsection{Analogy with beam splitters} An example of the sort of situation described above is for an ordinary non-polarizing beam splitter. Normally, two ports (for example ports $1$ and $2$ in Fig. \ref{bsfig}) are used as input, and two as output (ports $3$ and $4$). The beam splitter is then described by a unitary matrix, which can be taken to be \begin{equation}U^{(2)}_{BS}={1\over \sqrt{2}}\left(\begin{array}{cc}1 & i \\ i & 1\end{array} \right) .\end{equation} The two columns represent the two input directions, while the two outputs are represented by the rows. But if the beam splitter is connected in a network, where photons could be coming from any direction, any of the four ports could be used for either input or output at any given moment, so the matrix must have four input columns and four output rows and is of the form given in Eq. \ref{Hreversed}:
\begin{equation}U^{(4)}_{BS}={1\over \sqrt{2}}\left(\begin{array}{cccc} 0 & 0 & 1 & -i \\ 0 & 0 & -i & 1\\ 1 & i & 0 & 0\\  i & 1 & 0 & 0  \end{array} \right) =\left( \begin{array}{cc}0& U^{(2)\dagger}_{BS}\\ U^{(2)}_{BS} & 0\end{array}\right).\end{equation} It is easily verified in this case that $e^{-{{i\pi}\over 2}U^{(4)}_{BS}}$ is proportional to $U^{(4)}_{BS}$, so that a matrix of this form can serve as both Hamiltonian and evolution operator.

\begin{figure}[t!]
\centering
\includegraphics[totalheight=1.2in]{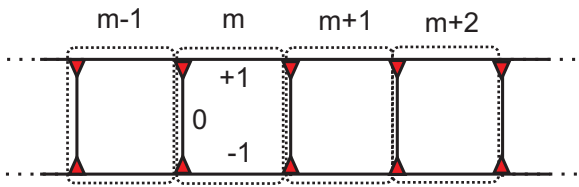}
\caption{(Color online) A system that can support three or six internal states per lattice site. $m$ labels the site, and three edges within the site are labeled $0,\pm1$. If propagation direction of the light is included, then this doubles the number of states per site.}
\label{threestatefig}
\end{figure}

\subsection{A discrete time-reversible system} Now consider a collection of three-ports connected to form the chain shown in Fig. \ref{threestatefig}. The chain is
divided again up into a string of lattice sites, indicated in the figure by the dashed boxes. Integer $m$ is used to label the lattice sites (i.e the horizontal
position), with $m$ increasing from left to right. The states within each site are labeled by an additional integer from among $\left\{ -1,0,+1 \right\}$: $+1$
means the photon is on the upper horizontal branch, $-1$ means the lower horizontal branch, and $0$ is the vertical branch. Further, we label the direction the
photon is moving at each moment according to the symbols $L$ or $R$. $L$ corresponds to leftward motion on a horizontal branch or upward on a vertical branch.
$R$ represents rightward or downward motion.  Therefore, the states are denoted by
\begin{equation}|m,j,D\rangle ,\label{states}\end{equation} where $m=1,\dots ,N$ for a lattice containing $N$ sites, $j\in \left\{0,-1,1\right\}$, $D\in \left\{ L,R\right\}$.

The input and detection considerations are similar to those of the last section. In the current case, since the evolution is left/right symmetric, it may be desired to couple the input state into the middle of the system (by means similar to that of Fig. \ref{compactfig}(b)), and make measurements at both ends of the chain.

The Hamiltonian at each cell should be of the form of Eq. \ref{Hreversed}, plus possible terms in the diagonal blocks due to inter-cell transitions. It is
straightforward to use the local transition matrix $U$ for the individual multiports to find the global transition amplitudes of the full system between the
states of Eq. \ref{states}. This gives the transition operator $U_c$ for the full chain. The resulting Hamiltonian $\hat H_c^\prime $
is:
\begin{widetext}
\begin{eqnarray}& & \hat H_c^\prime = {1\over 3}  \sum_{m=1}^N \Bigg\{ \sum_{j=-1}^1 \Big[ |m,j,R\rangle \langle m,j,L| + |m,j,L\rangle \langle m,j,R|\Big] 
\\ & &
-2\Big[ |m,1,R\rangle \langle m,0,L| +|m,0,R\rangle \langle m,1,L|  +   |m,-1,R\rangle \langle m,0,R| +|m,0,L\rangle \langle m,-1,L|
 \nonumber \\ & &
+|m-1,1,L\rangle \langle m,0,L| +|m-1,-1,L\rangle \langle m,0,R|  +  |m-1,1,L\rangle \langle m,1,L| +|m-1,-1,L\rangle \langle m,-1,L| 
 \nonumber \\ & &
+ |m+1,1,R\rangle \langle m,1,R| +|m+1,0,R\rangle \langle m,1,R|  +  |m+1,-1,R\rangle \langle m,-1,R| +|m+1,0,L\rangle \langle m,-1,R|  \Big] \Bigg\}  \nonumber
\end{eqnarray}\end{widetext}


$\hat H_c^\prime$ represents a system with one spatial dimension and a six-dimensional ``internal'' space. Experimentally, measuring the photon direction along with its position adds complications, so here we simplify matters by reducing to a three-dimensional internal space. If propagation direction is never measured, then the Hamiltonian can be projected onto the diagonal subspace of left- and right-moving modes. If the projection operator is $P$, then the resulting operator $\hat H_c=P^\dagger \hat H_c^\prime P$ becomes
\begin{widetext}\begin{eqnarray} \hat H_c &=& {1\over 3} \sum_m \left\{ \sum_{j=\left\{0,\pm 1\right\}} |m,j\rangle \langle m,j| \right. \label{Hc}
-\sqrt{2} \sum_{j=\left\{\pm 1\right\}}\Big[ |m,j\rangle \langle m,0|  +|m,0\rangle \langle m,j|\Big] \nonumber  \\ & & \qquad -\sqrt{2} \sum_{j=\left\{\pm 1\right\}}\Big[ |m-1,j\rangle \langle m,j|+|m+1,j\rangle \langle m,j| 
+|m-1,0\rangle \langle m,j|+|m-1,j\rangle \langle m,0| \Big] \Bigg\} .\nonumber\end{eqnarray}\end{widetext}

\begin{figure}
\centering
\includegraphics[totalheight=1.8in]{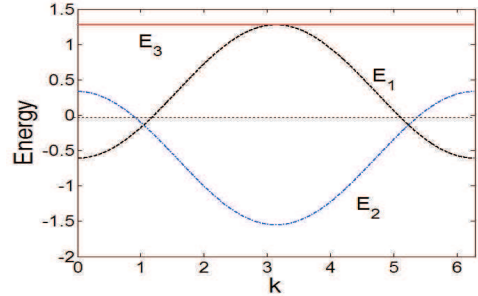}
\caption{(Color online) Energy level diagram of the three state system of Fig. \ref{threestatefig}. $E_3$ is constant, while $E_1$ and $E_2$ oscillate with $k$.  Note that  $E_1$ has no gap with $E_2$ or $E_3$, but that $E_2$ and $E_3$ are always separated by a gap.  }
\label{chain3fig}
\end{figure}

Going to momentum space, define \begin{equation}|m,j\rangle ={1\over \sqrt{N}}\sum_ke^{-imk}|k,j\rangle .\end{equation} The momentum space Hamiltonian then becomes \begin{widetext}\begin{eqnarray}\hat H_c&=& -{i\over 3} \sum_k \left\{  |k,0\rangle\langle k,0| + \left(  |k,1\rangle\langle k,1|  + |k,-1\rangle\langle k,-1|  \right)\left( 1-2\sqrt{2}\cos k\right) \right.  \\
\qquad &  & \left. -2\sqrt{2}\cos \left({k\over 2}\right) \left[ e^{ik/2}  \left(  |k,1\rangle  + |k,-1\rangle  \right) \langle k,0|
+ e^{-ik/2}  |k,0\rangle \left(  |k,1\rangle  + |k,-1\rangle  \right)
\right] \right\} \nonumber \\
&=& -{i\over 3}\sum_k |k\rangle \langle k| \left( \begin{array}{ccc} 1-2\sqrt{2} \cos k & -2\sqrt{2} e^{ik/2} \cos{k\over 2} & 0\\
-2\sqrt{2} e^{-ik/2} \cos{k\over 2} & 1 &  -2\sqrt{2} e^{-ik/2} \cos{k\over 2}\\
0 &  -2\sqrt{2} e^{ik/2} \cos{k\over 2} & 1-2\sqrt{2} \cos k
\end{array}\right) . \end{eqnarray}\end{widetext} The matrix is written here in the basis of internal states, $(+1,0,-1)$.  It can be diagonalized in order to find the eigenstates, leading to a set of three eigenvectors for each fixed $k$ value, given (up to normalization and phase) by \begin{eqnarray}& & |\psi_1\rangle =\left( \begin{array}{c} -1\\ 0\\ 1  \end{array}\right), \qquad |\psi_2\rangle =\left( \begin{array}{c} 1\\ -2\sqrt{2}\\ 1  \end{array}\right), \\ & &  |\psi_3\rangle =\left( \begin{array}{c}-1\\ 2\sqrt{2}(1+\cos k)\\ 1  \end{array}\right),\end{eqnarray}
with respective energy eigenvalues \begin{eqnarray} E_1&=& {1\over 3}\left( 1-2\sqrt{2}\cos k \right) \\
E_2&=& {1\over 3}\left( 1-2\sqrt{2}\left( 1-\cos k\right) \right)  \\
E_3&=& {1\over 3}\left( 1+2\sqrt{2} \right) . \end{eqnarray}
These are plotted in Fig. \ref{chain3fig}. Notice that the energy eigenvalues always satisfy \begin{equation}E_1+E_2+E_3 =1,\end{equation} for all $k$. In all,
there are $3N$ eigenvectors, corresponding to the $N$ momentum values ($k={{2\pi n}\over N}$ for $n=1,\dots N$) and the three values of the internal index
($0,\pm 1$). As in the last section, the constant  terms on the diagonal contribute a constant ($x$- and $k$-independent) background term, in this case of energy
${1\over 3}$.

\subsection{Energy band structure} For each $k$, there are three energy levels. They are nondegenerate in general, except at isolated $k$ values where the levels cross
or become tangent to each other. Note also that there is always an energy gap between $E_2$ and $E_3$, but that $E_1$ has $k$ values where its gap with the other
two states closes. Levels $1$ and $3$ meet ($E_1=E_3$) at the value $k=\pi$, while $E_1=E_2$ at the values $k=\pm {\pi\over 3}$. Thus, at those momenta there can
be transitions from eigenstate $|\psi_1\rangle $ to the other two states, $|\psi_2\rangle $ and $|\psi_3\rangle $, if the system is perturbed. However,
$|\psi_2\rangle $ and $|\psi_3\rangle $ never transition directly to each other; they can only go via $|\psi_1\rangle$ as an intermediate state; moreover, such a
transition would be indirect in the terminology of solid state physics, since it requires a change of $k$ as well. Because transitions between $|\psi_2\rangle$
and $|\psi_3\rangle$ are always mediated by passage through $|\psi_1\rangle $, $|\psi_1\rangle$ plays a role similar to that of a photon or other intermediate
gauge boson in particle physics. A more sophistication simulation setup along these lines might therefore eventually open the way for optical simulation of gauge
theory models such as quantum chromodynamics, with the role of nonlinear interactions being simulated via interference effects.

As mentioned in Section \ref{discsection}, the Hamiltonian can be written in the form \begin{equation}\hat H_c= d_0 I +\sum_{j=1^8}d_j\Lambda_j = d_0 I +\bm d
\cdot \bm \Lambda,\end{equation} where the Gell-Mann $\Lambda_j$ matrices play the same role for the algebra $su(3)$ that the Pauli matrices play for $su(2)$. As
the parameters are varied, the vector $d=(d_0,d_1,\dots,d_8)$ does not explore the full nine-dimensional space, but remains contained in a 7-dimensional subspace
spanned by the set $\left\{I,\Lambda_1,\Lambda_2,\Lambda_3,\Lambda_6,\Lambda_7,\Lambda_8\right\}$. The explicit form of the Hamiltonian in terms of the
$\Lambda_j$ may be found in the Appendix. The paths on this large subspace have enough room to avoid topological obstructions as the system parameters are
varied, so that any of the possible paths can be smoothly deformed into any other as the parameters are continuously changed. The system is therefore always
topologically trivial. It remains to be investigated as to whether varying the parameters of the three-port may allow simulation of more general $SU(3)$
Hamiltonians.

\section{Conclusions}

Through the examples discussed in the previous sections, it has been shown that the behavior of Hamiltonian systems with both spatial and internal degrees of
freedom can be simulated using one-dimensional chains of directionally-unbiased three-ports. A number of obvious generalizations exist, such as using $n$-ports
with higher $n$, connecting the $n$-ports into two- and three-dimensional arrays with different connection topologies, or varying the multiport parameters. In this way, a variety of Hamiltonians involving nearest neighbor couplings
can be implemented. By changing the phase shifts at the multiport vertices, the relative strengths of the interactions between different nearest-neighbor pairs
can be altered, allowing the simulation of spatially-varying potentials. Additional degrees of freedom that can be used to increase the complexity of the simulated systems include phase shifts on the edges of the multiports or between multiports, and varying the transmission profile of the beam splitters within the multiports. The results of the previous sections provide hints that such generalized versions of
this approach may be useful for simulating the types of Hamiltonians that appear in solid state physics and particle physics, possibly including strongly
interacting Hamiltonians for which perturbative methods break down.

One generalization of special interest is that the simulating system can be built out of unit cells that are combinations of several multiports, possibly with additional phase shifts or other effects added on the connections between them. By periodically altering the parameters of these complex unit cells, the behavior of particles in periodic crystal lattices can be simulated. In particular, it will be shown elsewhere \cite{simtop} that by this means the Su-Schreiffer-Heeger Hamiltonian, which supports phases of nonzero winding number and topologically-protected boundary states, can be simulated.

In general, it seems to be possible to simulate arbitrary physical systems with nearest neighbor interactions by varying the parameters appropriately to adjust the allowed energy levels. Similarly, varying these parameters spatially along the chain can simulate interactions with arbitrary external (discrete-space) potentials.  Variation of the phases allows for the positions of the energy levels to be varied in a controllable manner. For example, the setup in Section \ref{3pointsection} has two degenerate levels. The degeneracy is due to the equality of all the vertex phases and can be lifted by changing those phases; for some parameter ranges, if two phases are held constant then the energy-level splitting varies approximately linearly with the other phase, allowing easy control of the system's behavior. This could be used for example to simulate the behavior of systems in magnetic fields, with the difference between two of the vertex phases playing the role of the field, or to simulate three- or four-level atomic systems.

Such generalizations hold promise for simulating a large range of effects, and possibly carrying out complex
high-dimensional computations of the properties of such discrete-time systems.  It should be emphasized again that all of these
simulations are implemented by means of local linear optics effects only. In addition, all of the elements used can be placed on-chip, which increases stability and allows large networks to be built up in a highly scalable manner. As a result, this approach seems especially promising for the simulations of complex physical systems by relatively
simple means.

\section*{Appendix}

The Hermitian Gell-Mann matrices form a basis for the $8$-dimensional \emph{algebra} of $su(3)$ in the same manner that the Pauli matrices form a basis of the
three-dimensional
$su(2)$.  Explicitly, these matrices are given by \cite{arfken} \begin{eqnarray}\Lambda_1 =\left(\begin{array}{ccc}  0 & 1 & 0\\ 1 & 0 & 0 \\ 0 & 0 & 0 \end{array} \right) \quad & & \quad   \Lambda_2 =\left(\begin{array}{ccc}  0 & -i & 0\\ i & 0 & 0 \\ 0 & 0 & 0 \end{array} \right) \\
\Lambda_3 =\left(\begin{array}{ccc}  1 & 0 & 0\\ 0 & -1 & 0 \\ 0 & 0 & 0 \end{array} \right) \quad & & \quad   \Lambda_4 =\left(\begin{array}{ccc}  0 & 0 & 1\\ 0 & 0 & 0 \\ 1 & 0 & 0 \end{array} \right) \\
\Lambda_5 =\left(\begin{array}{ccc}  0 & 0 &  -i\\ 0 & 0 & 0 \\ i & 0 & 0 \end{array} \right) \quad & & \quad   \Lambda_6 =\left(\begin{array}{ccc}  0 & 0 & 0\\ 0 & 0 & 1 \\ 0 & 1 & 0 \end{array} \right) \\
\Lambda_7 =\left(\begin{array}{ccc}  0 & 0 &  0\\ 0 & 0 & i \\ 0 & i & 0 \end{array} \right) \quad & & \quad   \Lambda_8 ={1\over \sqrt{3}}\left(\begin{array}{ccc}  1 & 0 & 0\\ 0 & 1 & 0 \\ 0 & 0 & -2 \end{array} \right) \end{eqnarray}
The corresponding $SU(3)$ \emph{group} generated by exponentiating elements of the algebra is spanned by these matrices together with the identity matrix.

In terms of the group basis, the Hamiltonian of Eq. \ref{Hc} can be written as \begin{eqnarray}\hat H_c &=& {1\over 3}\left\{ \left( 1-{{4\sqrt{2}}\over 3}\cos k
\right) I +\sqrt{2}\Bigg[ {{ \Lambda_8}\over \sqrt{3}} -\Lambda_3 \right.  \\ & & \left.-\left( 1+\cos k\right) \left( \Lambda_1 +\Lambda_6\right) +\sin k \left(
\Lambda_2 -\Lambda_7\right)\Bigg]\right\},\nonumber \end{eqnarray} which lives on the seven-dimensional subspace of the full nine-dimensional group space spanned
by the basis elements $\left\{I,\Lambda_1,\Lambda_2,\Lambda_3,\Lambda_6,\Lambda_7,\Lambda_8\right\}$. Any closed path in this space is contractible: it can be
shrunken to a point while avoiding the missing zero-energy point at the origin. As a result all Hamiltonians in the space are topologically trivial.

\section*{Acknowledgements} This research was supported by the National Science Foundation EFRI-ACQUIRE Grant No. ECCS-1640968, NSF Grant No. ECCS- 1309209, and by the Northrop
Grumman NG Next.


\begin{thebibliography}{99}

\bibitem{feyn} R. P. Feynman, Int. J.  Theor. Phys.  \textbf{21}, 467 (1982)

\bibitem{aspuru} A. Aspuru-Guzik, and P. Walther, Nat. Phys. \textbf{8}, 285 (2012)

\bibitem{johnson} T. H. Johnson, S. R. Clark, and D. Jaksch, EPJ Quantum Technology \textbf{1}, 10 (2014)

\bibitem{threeport} 
D. S. Simon, C. A. Fitzpatrick and A. V. Sergienko,
    Phys. Rev.  A \textbf{93}, 043845 (2016).

\bibitem{hbf} M. Hillery, J. Bergou, E. Feldman, Phy. Rev. A \textbf{68}, 032314, (2003)

\bibitem{fh1} E. Feldman, M. Hillery, Phy. Lett. A \textbf{324}, 277 (2004)

\bibitem{fh2} E. Feldman, M. Hillery, Cont. Math. \textbf{381}, 71 (2005) 

\bibitem{fh3} E. Feldman, M. Hillery, J. Phys. A: Math. Theor. \textbf{40},  11343 (2007)

\bibitem{tara} B. Tarasinski, J. K. Asb\'oth, J. P. Dahlhaus, Phys. Rev. A {\bf 89}, 042327 (2014)

\bibitem{simtop} D. S. Simon, C. A. Fitzpatrick and A. V. Sergienko, submitted for publication (2017)

\bibitem{carn} I. Carneiro, M. Loo, X. Xu, M. Girerd, V. Kendon, P. L. Knight, New J. Phys. \textbf{7},  156 (2005)

\bibitem{kitagawa} T. Kitagawa, Quant. Inf. Proc. \textbf{11}, 1107 (2012)
%
%
%
%

\bibitem{asboth} J. K. Asb\'oth, L. Oroszl\'any, A. P\'alyi, \emph{A Short Course on Topological Insulators}, Springer, Berlin (2016)

\bibitem{hasan} M. Z. Hasan, C. L. Kane, Rev. Mod. Phys. \textbf{82}, 3045 (2010)

\bibitem{arfken} G. B. Arfken, H. J. Weber, F. E. Harris, \emph{ Mathematical Methods for Physicists}, 7th ed., Academic Press, London (2012)

\end{thebibliography}
\end{document}